\documentclass[aps,prc,twocolumn,groupedaddress,showpacs]{revtex4}

\usepackage{graphicx}

\begin{document}
\title{Evidence for  a dynamically refracted primary bow  in 
weakly bound \\$^9$Be    rainbow  scattering  from $^{16}$O 
   }

\author{
S. Ohkubo$^{1}$ and  Y. Hirabayashi$^2$  
 }
\affiliation{$^1$Research Center for Nuclear Physics, Osaka University, 
Ibaraki, Osaka 567-0047, Japan }
\affiliation{$^2$Information Initiative Center,
Hokkaido University, Sapporo 060-0811, Japan}

\date{\today}

\begin{abstract}
 We present for the first time
  evidence for the existence  of a dynamically refracted primary bow 
for   $^{9}$Be+$^{16}$O scattering.
 This is demonstrated    through the use of  
  coupled channel calculations  with an extended double folding
 potential derived from the density-dependent effective two-body force and   
precise microscopic cluster wave functions for $^{9}$Be. The calculations reproduce  the
experimental Airy structure in $^{9}$Be+$^{16}$O scattering well.
It is found that   coupling of   a  weakly bound  $^{9}$Be nucleus to excited states  plays
 the role  of a booster lens, dynamically enhancing the  refraction  over the {\it static} 
 refraction  due to  the Luneburg lens  mean field potential  between the ground states of $^{9}$Be and
 $^{16}$O.
 \end{abstract}

\pacs{25.70.Bc,24.10.Eq,24.10.Ht}
\maketitle

\section{INTRODUCTION}
\par
A nuclear rainbow has been understood to be caused by  refraction  within the static 
Luneburg lens mean field potential  during elastic scattering
 \cite{Ford1959,Brink1985, Michel2002}. 
The existence of the nuclear rainbows has been  confirmed experimentally
under relatively weak absorption for many systems such as $\alpha$+$^{40}$Ca, $\alpha$+$^{16}$O,
$^{16}$O+$^{16}$O, $^{16}$O+$^{12}$C, and   $^{12}$C+$^{12}$C \cite{Goldberg1972,Khoa2007}.
Very recent studies have shown 
the existence of a secondary bow   in nuclear  rainbow scattering 
 involving $^{12}$C, which is not caused by a {\it static} Luneburg lens mean field potential
 but  is 
generated by  a {\it dynamical} coupling to the excited states \cite{Ohkubo2014,Ohkubo2015B}.
The specific structure of a strongly deformed $\alpha$ cluster nucleus $^{12}$C is
 related to the dynamical generation  of a secondary  bow.
It is therefore interesting and intriguing to investigate whether  a 
 dynamically refracted rainbow can exist in other systems for which coupling to the excited states
 is important.
 
\par
 The $^{9}$Be nucleus  is  strongly deformed with a quadrupole deformation parameter 
$\beta_2$=1.4 \cite{Tanaka1978}, which is larger than that of $^{12}$C, $\beta_2$=-0.40
 \cite{Yasue1983} and  is weakly bound  with 
the threshold energy 1.57 MeV, 1.67 MeV, and 2.47 MeV for the $\alpha$+$\alpha$+$n$, 
 $^8$Be+$n$, and  $\alpha$+$^5$He decays,
respectively.
There have been extensive studies of  the effect of  breakup channels of weakly bound nuclei   on the polarization
 potential    
\cite{Thompson1981,Nagarajan1982,Gomez1985,Sakuragi1983,Sakuragi1987,Hirabayashi1989}.
 No attention has been paid to nuclear rainbows.

\par
The systematic study of a weakly bound  $^6$Li scattering from  $^{12}$C and $^{16}$O
 over a wide range of energies using a phenomenological potential has 
shown \cite{Michel2005,Michel2007} the existence of  a  nuclear rainbow  and 
  Airy structure  in the angular distributions.
This shows that in contrast to a naive    strong absorption picture 
absorption of scattering involving weakly bound nuclei 
 is not complete.
 As for $^{9}$Be,  the experiment  of $^9$Be+$^{16}$O scattering in the rainbow energy region
was performed   at $E (^9$Be)=157.7 MeV  
\cite{Satchler1983A,Fulmer1984}.  Satchler {\it et al.} \cite{Satchler1983A}
 interpreted that  the observed
  rainbow-like behavior  of the angular distribution is  a nuclear rainbow ``ghost".
Khoa \cite{Khoa1988}  reproduced   the angular distribution of the   nuclear rainbow
 ``ghost"  using  double folding model calculations 
with the   M3Y force well by taking into account the  finite-range exchange effect. 
Recently Glukhov {\it et al.} \cite{Glukhov2010}   measured   $^{16}$O+$^9$Be scattering 
at $E(^{16}$O)=132 MeV  and reported   the existence of an Airy minimum 
in the phenomenological optical model analysis.

\par
It is  important   to definitively confirm   the existence of a nuclear  rainbow
 theoretically  and also to investigate   how 
  a rainbow is  generated in weakly bound 
 $^9$Be scattering. The relevance of   the breakup    to the 
emergence  of a nuclear rainbow is especially important.
 Also it is intriguing  to investigate whether a double folding model derived from 
 the density-dependent force, 
 which has been successful  in many   systems involving non-weakly bound nuclei such as 
 $\alpha$ particle, $^{16}$O,  $^{12}$C, and $^{14}$C 
\cite{Khoa2007,Abele1993,Atzrott1996,Hirabayashi2013,Ohkubo2015},
  can describe   the nuclear rainbow phenomenon involving a weakly bound 
 $^9$Be nucleus well.

\par
The purpose of this paper is to present for the first time  evidence for  the existence 
of a dynamically  refracted primary bow by studying  the mechanism of generation of
 a nuclear rainbow in scattering of a weakly bound   $^9$Be nucleus  from $^{16}$O.
 Coupled channel calculations with an  extended   
double folding model potential derived from the precise  wave functions of $^9$Be 
and the density-dependent effective force are performed. It is shown that refraction 
  is   boosted  by coupling to the  excited states. This boosted dynamical refraction
in addition to the static refraction in the  mean field potential manifests itself in
 the observation of a  primary bow in nature.  

\begin{figure}[t]
\includegraphics[keepaspectratio,width=7.8cm] {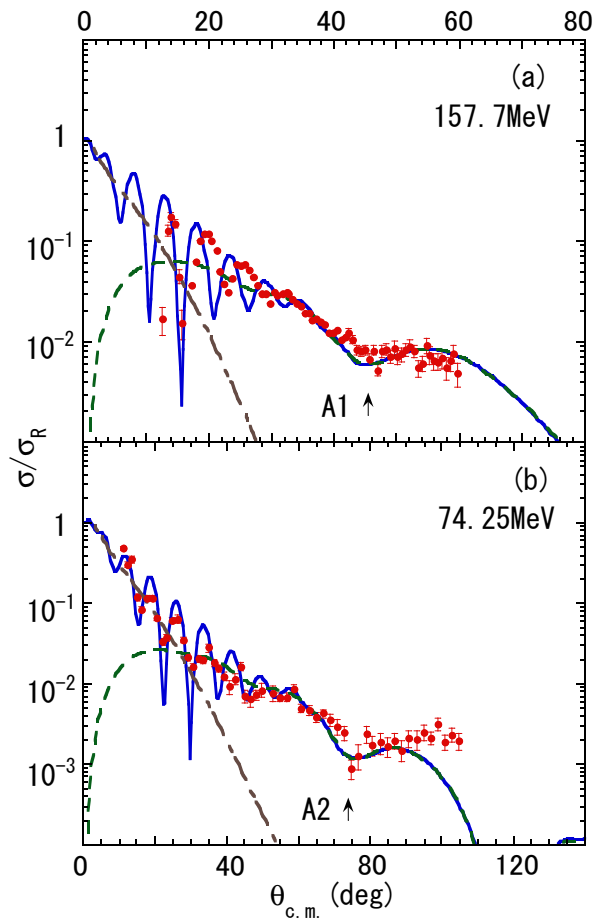}
 \protect\caption{\label{fig.1} {(Color online) 
The angular distributions of $^{9}$Be+$^{16}$O  scattering at 
(a)  $E (^9$Be)=157.7 MeV  and 
(b) 74.25 MeV ($E(^{16}$O)=132 MeV)   
 calculated in a single channel (solid lines)  are compared with the 
experimental data (points) \cite{Satchler1983A,Fulmer1984,Glukhov2010}.
The calculated cross sections are  decomposed into the farside
component (dashed lines) and the nearside component (dash-dotted lines).
}
}
\end{figure}

\section{EXTENDED DOUBLE FOLDING MODEL}
\par
We study     $^{9}$Be+$^{16}$O scattering using the coupled channel (CC) method  
with an extended double folding 
(EDF) model that describes all the diagonal and off-diagonal coupling potentials 
derived from a density-dependent   nucleon-nucleon force and the precise  microscopic  
  wave functions for $^{9}$Be  and $^{16}$O.
  The diagonal and coupling potentials 
for the $^{9}$Be+$^{16}$O system are calculated using the EDF  model
\begin{eqnarray}
\lefteqn{V_{ij}({\mathbf R}) =
\int \rho_{ij}^{\rm (^{9}Be)} ({\bf r}_{1})\;
     \rho^{\rm (^{16}O)} ({\bf r}_{2})} \nonumber\\
&& \times v_{\it NN} (E,\rho,{\bf r}_{1} + {\mathbf R} - {\bf r}_{2})\;
{\it d}{\bf r}_{1} {\it d}{\bf r}_{2} ,
\end{eqnarray}
\noindent where 
$\rho_{ij}^{\rm (^{9}Be)} ({\bf r})$ represents the diagonal ($i=j$) or transition ($i\neq j$)
 nucleon density of $^{9}$Be which is calculated using the microscopic  molecular  model
in the generator coordinate  method \cite{Okabe1979}. This model reproduces  the 
energy  spectra, electromagnetic properties, charge form factors, neutron and 
$\alpha$ decay widths of  $^{9}$Be  well   \cite{Okabe1979}. 
In the coupled channel calculations we include the ground band states of 
$^{9}$Be, $3/2^-$ (0.0 MeV), $5/2^-$ (2.43 MeV) and $7/2^-$ (6.38 MeV) \cite{Tilley2004}. 
Other states (for example, $1/2^+$ (1.68 MeV), $1/2^-$ (2.78 MeV) and $5/2^+$ (3.05 MeV)) 
 are found not to contribute 
significantly in the present coupled channel calculations.   
$\rho^{\rm (^{16}O)} ({\bf r})$ is the  nucleon 
 density of  $^{16}$O 
  taken from   Ref.\cite{Okabe1995}. For the  effective interaction   $v_{\rm NN}$     we use  
 the Density Dependent Michigan 3
range Yukawa-Finite Range (DDM3Y-FR) interaction \cite{Kobos1982}, which takes into account the
finite-range nucleon  exchange effect \cite{Khoa1994}.
 We introduce the normalization factor  $N_R$ \cite{Brandan1997,Khoa2001} for 
 the real double folding potential. 
An imaginary potential with a  Woods-Saxon volume-type (nondeformed) form factor is 
introduced   phenomenologically to take into account the effect
of absorption due to other channels.  
   A complex coupling, which is often used but has no rigorous theoretical justification
 especially for a composite projectile   \cite{Satchler1983}, is not introduced 
because without it 
 the present EDF model  successfully reproduced  many  rainbow scattering data 
systematically over a wide range of incident energies 
\cite{Ohkubo2004A,Ohkubo2007,Hamada2013,Hirabayashi2013,Ohkubo2014,Ohkubo2014B,Ohkubo2014C,Mackintosh2015,Ohkubo2015,Ohkubo2015B}.

\begin{table}[t]
\begin{center}
\caption{ \label{Table I}
The normalization factor $N_R$,  volume integral per nucleon pair $J_V$  of the 
 the ground state diagonal potential (in units of MeVfm$^3$),  and the   imaginary potential
 parameters used in the
 single channel double folding calculations  and coupled channel calculations   with EDF 
in Fig.~1 and Fig.~2.  
}
\begin{tabular}{clccccclccc}
 \hline
  \hline
$E(^{9}{\rm Be})$ & $E(^{16}{\rm O})$ & $N_R$& $J_V$  & $W$    & $a$ &   & $N_R$ &$J_V$
 &$W$  &$a$  \\
 &  &    \multicolumn{4}{c}{ (single channel cal)}   &  &
\multicolumn{4}{c}{(coupled channel cal)}   \\           
 \cline{3-6}    \cline{8-11}
(MeV) & (MeV) & & &(MeV) &(fm)  &  & & & (MeV)  & (fm)    \\   
 \hline
 \hline
 74.25 &132 &1.1 &393 &    21    & 0.9&  &  1.03& 368 &  17 & 1.0   \\
 157.7& 280.4 & 1.1& 356 &  24      & 0.95& &  1.00 & 324 & 20  & 1.0   \\
 \hline                          
 \hline                          
\end{tabular}
\end{center}
\label{Table1}
\end{table}

\begin{figure}  [t]
\includegraphics[keepaspectratio,width=7.8cm] {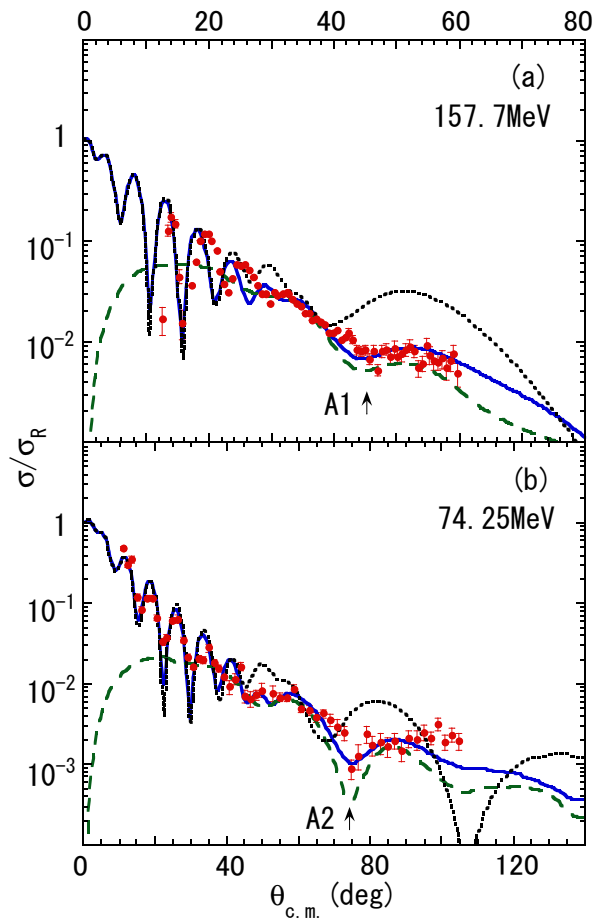}
 \protect\caption{\label{fig.2} {(Color online) 
 The angular distributions of $^{9}$Be+$^{16}$O  scattering at 
(a)  $E (^9$Be)=157.7 MeV  and (b)
 74.25 MeV ($E(^{16}$O)=132 MeV)  calculated using  the 
 coupled channel method  with coupling to    both the $5/2^-$ and $7/2^-$ states 
including  reorientations (solid lines) are compared with the 
experimental data (points) \cite{Satchler1983A,Fulmer1984,Glukhov2010}.
The  farside  component of the calculated cross sections is indicated   by the dashed lines.
The single channel calculations where the coupling is switched off are displayed for
 comparison with  dotted lines.
}
}
\end{figure}

 \begin{figure}[t]
\includegraphics[keepaspectratio,width=7.8cm] {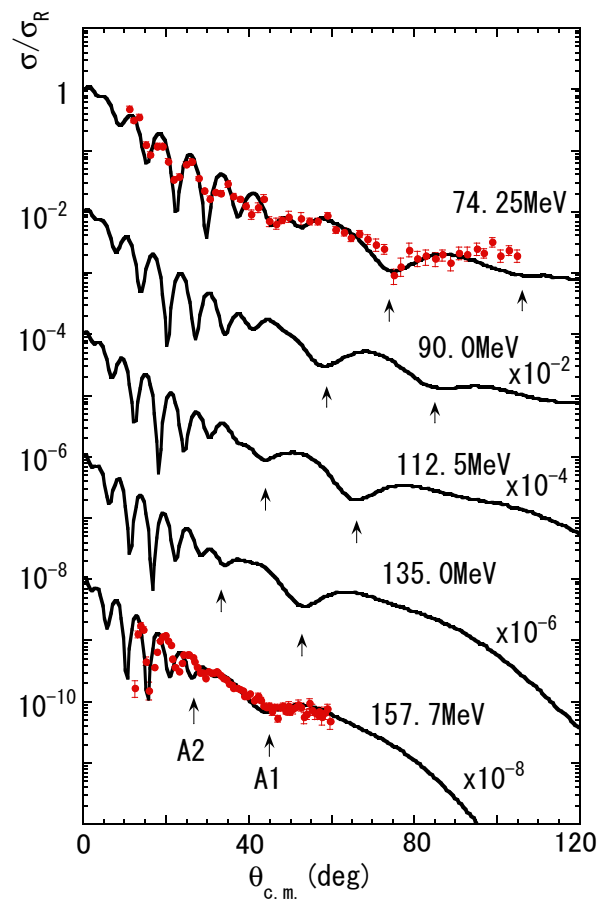}
\protect\caption{\label{fig.3} {(Color online) 
 The energy evolution of  the Airy minimum $A1$ and $A2$ of the Airy structure in 
the  angular  distributions between 
 $E (^9$Be)=157.7 MeV  and  74.25 MeV ($E(^{16}$O)=132 MeV), calculated using the 
CC method,  are displayed with solid lines. The experimental data 
\cite{Satchler1983A,Fulmer1984,Glukhov2010} are indicated by the points. }
}
\end{figure}

\section{ANALYSIS OF $^9$B\lowercase{e} + $^{12}$C SCATTERING}
\par
 We analyze  the angular distributions of $^{9}$Be+$^{16}$O  scattering
at  $E(^{9}$Be)=157.7 MeV ($E_{c.m.}$=100.9 MeV) \cite{Satchler1983A,Fulmer1984} and 
  $E(^{16}$O)=132 MeV ($E_{c.m.}$=47.5 MeV, $E(^{9}$Be)=74.25 MeV) \cite{Glukhov2010}
 in the rainbow energy region.
Hereafter, the incident energies are given in the frame of  $E(^{9}$Be).
In Fig.~1 the angular distributions calculated  in a single channel   are displayed
 in comparison with the experimental data. 
In the calculations the reorientation of the ground state is not included.
 The value of $N_R$ for the real potential was adjusted to 
fit the data.  For the imaginary potential a radius parameter    was fixed at   $R$=5.5 fm
 while  a strength  parameter of around  $W$=$20$ MeV and a diffuseness parameter of around
 $a=1.0$ fm were found to fit the data.   The values of  $N_R$  together with the volume
 integral per nucleon pair of the real potential, $J_V$, and potential parameters are
given in Table I.
The characteristic features of a nuclear  rainbow scattering in the  experimental angular
 distributions are  reproduced well by the calculations
by only slightly changing the value of $N_R$ from unity to $N_R$=1.1.
The calculated cross sections are decomposed into the farside and nearside contributions.
The angular distributions at the forward angles are Fraunhofer diffractive scattering,
 which are  sensitive only to  the surface region of the potential \cite{Hirabayashi1989}, 
and in the present calculations they are  caused by the interference between the nearside
 and farside contributions as seen in Fig.~1.
On the other hand, the angular  distributions in the intermediate angular region are 
dominated by only  farside refractive  scattering, which penetrates deep into the internal 
region of the potential.  Thus the  minima in the experimental angular distributions 
at $\theta=45^\circ$ in Fig.~1(a)  and  $78^\circ$  in Fig.~1 (b) are found to be  Airy 
minima of the nuclear primary rainbow.
  At 157.7 MeV in Fig.~1(a) the   calculation  reproduces  the Airy minimum 
  in agreement with  the  experimental  minimum. 
 The  Airy minimum at $\theta=45^\circ$ and the Airy maximum  at around
 $\theta=55^\circ$  are assigned    to be of first order, $A1$
because  beyond that there is a fall-off in    the  angular distribution,
 which is the appearance of the darkside of a nuclear rainbow.  Although the fall-off  has not been
measured in experiment, it is clear that the  minimum and  maximum are not the
 ``ghost" of  the nuclear rainbow but the real rainbow due to  refractive scattering. 
That this is really the nuclear rainbow can be further  confirmed in  the lower energy
 experimental data at 74.25 MeV in Fig.~1(b) by identifying the existence of 
  the higher order Airy minimum of the nuclear primary rainbow.
The calculation   reproduces  the characteristic features of the experimental 
angular distribution  with a minimum  at $78^\circ$, which 
 is a second order  Airy minimum $A2$ well. This  order 
 will be shown  without ambiguity  by investigating the energy  evolution of the angular
 position of  the Airy minimum, as  displayed in Fig.~3. 
Some enhancement of the experimental cross sections beyond $\theta=90^\circ$ at 74.25 MeV 
compared with  the calculation  may be due to effects other than refractive
 scattering such as exchange effects.

\par
In order to reveal the role of the  excited states on the emergence of the primary nuclear
  rainbow,  the angular distributions 
 of elastic $^{9}$Be+$^{16}$O  scattering calculated  using   the coupled channel method
 including  reorientations   are  displayed in Fig.~2.  
For the imaginary potential, the strength parameter $W$  is slightly  readjusted to decrease
  because of channel coupling while  the radius parameter was kept $R=5.5$ fm.
The values of $N_R$ needed are almost unity and slightly smaller than those 
 in the single channel
 calculations in Fig.~1.  The potential parameters used and the  values of the volume integral per
 nucleon pair of the double folding  potential, $J_V$,  are given in Table I.
The CC calculations  reproduce  the  Airy structure of the experimental angular 
distributions well.
In Fig.~2(a) the  minimum $A1$ at  $\theta=45^\circ$ is seen to be caused  by farside refractive
 scattering also in the coupled channel calculation. In the calculation without coupling 
(dotted lines),
 although  the Airy minimum $A1$ is seen,  it is  located at a considerably   smaller  angle
 $\theta=38^\circ$ in disagreement with the  experimental  $\theta=45^\circ$.
 This means the attraction is insufficient  without coupling to the excited states. 
 By introducing coupling to the excited states,  the  Airy minimum $A1$ moves backward 
in agreement with the experimental data as seen  in Fig.~2(a). 
In Fig.~2(b) the same situation is seen at the lower energy of  $E (^9$Be)=74.25 MeV.
 The experimental Airy minimum $A2$ at  $\theta=78^\circ$  is correctly reproduced by the 
CC calculation.  The dominance of the farside  contribution in the CC calculation 
in the intermediate angular region shows that this minimum is really an Airy minimum due
to  refractive scattering.  In the calculation without channel coupling (dotted lines),
 although the Airy
 minimum $A2$ is seen, it is located considerably farther forward, at an  angle $\theta=65^\circ$.
This means that attraction is lacking considerably if the coupling to the excited states are
absent. 
Thus it is found that   coupling  to the  excited  $5/2^-$ and $7/2^-$ states 
plays the role of inducing additional attraction. Namely, the coupling  plays the role of
 a booster second lens causing additional refraction over that due to the static Luneburg lens 
mean field  potential caused by the ground state. Thus the emergence of the primary nuclear  
rainbow for this system is found to be realized in nature by the dynamically boosted refraction
 due to the coupling to the  excited states.

\begin{figure}[t]
\includegraphics[keepaspectratio,width=7.8cm] {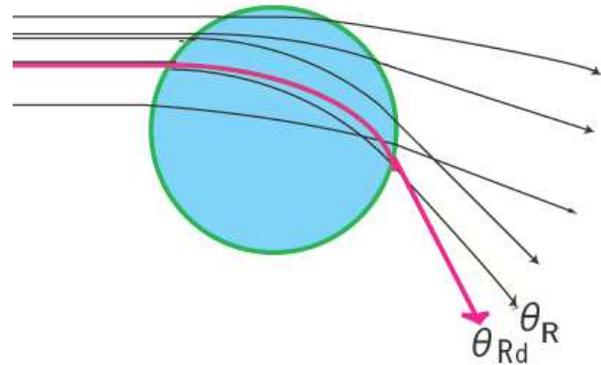}
 \protect\caption{\label{fig.5} {(Color online) 
  Illustrative refractive trajectories (solid lines)  in  nuclear   rainbow scattering. 
 The refracted angle $\theta_R$ is a
 rainbow angle  for the  primary nuclear rainbow caused by the 
 mean field optical potential (Luneburg lens \cite{Michel2002}) of the nucleus (indicated by a circle)
without coupling to the excited states.  The  angular region
 $\theta$$\leq $$\theta_R$ is the  bright side of the primary nuclear rainbow and 
$\theta$$>$$\theta_R$ is the darkside. By coupling to the excited states of $^9$Be 
the refraction is dynamically enhanced and the rainbow angle is increased
to   $\theta_{Rd}$.  
}
}
\end{figure}

\par
 In Fig.~3 the energy evolution   of the Airy structure for $^{9}$Be+$^{16}$O  scattering
 is shown, by displaying  the angular distributions for a range of energies 
   between   $E (^9$Be)=157.7  and  74.25 MeV. The angular distributions are calculated in the three  coupled channel
 calculations using the interpolated potential parameters. 
The angular position of the Airy minimum moves backward as the energy decreases.
The $A1$ located at $\theta=45^\circ$ at 157.7 MeV  moves backward to $105^\circ$ at
74.25 MeV, whose existence could  be confirmed by    measurement at  larger angles.
 On the other hand, the $A2$ that is located at forward angles at 157.7 MeV, and which is 
difficult to see in the experimental data due to being masked by the Fraunhofer diffraction, 
 develops moving backward as the incident energy decreases. At 74.25 MeV this Airy minimum
   is clearly identified as that of order two, $A2$, at $\theta=78^\circ$ in the experimental
angular distribution. Thus the existence of a nuclear rainbow together with the higher order
 Airy structure is  confirmed for  scattering involving a weakly bound  $^9$Be nucleus.

\begin{figure}  [t]
\includegraphics[keepaspectratio,width=7.8cm] {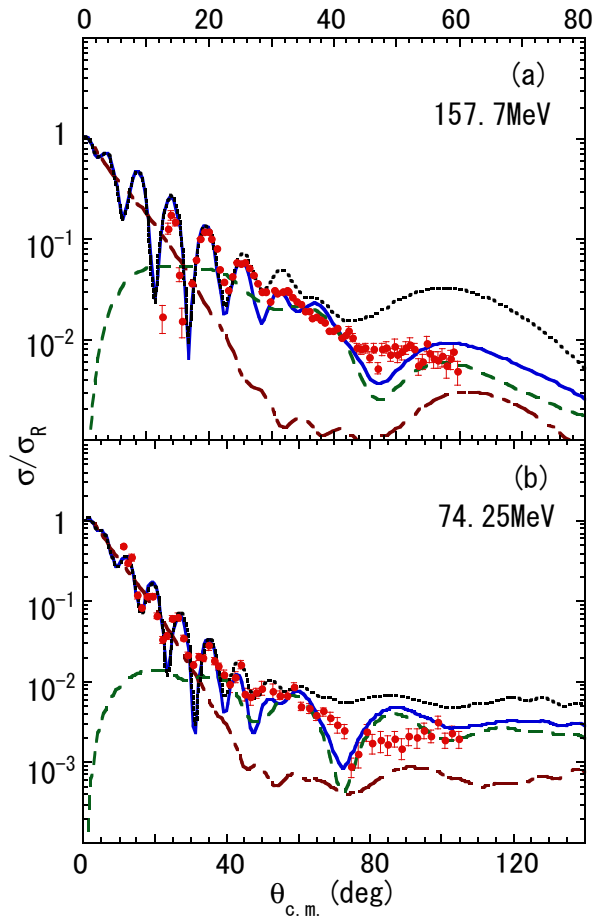}
 \protect\caption{\label{fig.5} {(Color online) 
Same as Fig.~2 but for the calculations with the  M3Y force.
 The angular distributions of $^{9}$Be+$^{16}$O  scattering at 
(a)  $E (^9$Be)=157.7 MeV  and (b)
 74.25 MeV ($E(^{16}$O)=132 MeV)  calculated using  the 
 coupled channel method  with coupling to    both the $5/2^-$ and $7/2^-$ states 
including  reorientations (solid lines) are compared with the 
experimental data (points) \cite{Satchler1983A,Fulmer1984,Glukhov2010}.
The  farside and nearside component of  cross sections calculated using the
 CC method are indicated with  dashed  lines and dashed-dotted lines, respectively.
The single channel calculations where the coupling is switched off are displayed for
 comparison with  dotted lines.
}
}
\end{figure}

\section{PRIMARY BOW AND BOOSTER LENS} 
\par
In Fig.~4 an illustrative figure of the enhanced refraction, boosted dynamically by
   coupling to excited states of $^{9}$Be, is displayed. The incident projectile
is refracted by the static potential due to the ground state of the target nucleus $^{16}$O.
The largest refractive angle (rainbow angle, deflection angle) in the static potential
 is indicated by $\theta_R$. Because the  projectile is easily excited, the strong coupling
 to the excited   states of the ground band  above the threshold energy for breakup causes
 additional refraction. Thus the largest refractive angle is increased to 
$\theta_{Rd}$.  
The excited states play the role of a  dynamical booster lens, enhancing the 
refraction  over the static refraction caused by the Luneburg lens  mean field ground state
 potential. The observed   rainbow may be called a dynamically refracted rainbow 
or a dynamically boosted rainbow because it is not realized in nature without this boosting.
From  Table I, we can determine quantitatively the induced attraction  by considering 
 changes of the value of
$J_V$:  $\Delta$$J_V$=25 MeVfm$^3$    at 157.7 MeV   and 
$\Delta$$J_V$=32 MeVfm$^3$  at 74.25 MeV. 
About 10\% of the  refraction is due to dynamical effects.

In order to confirm  the dynamical refraction  we have also calculated 
by using other effective forces.  Hitherto we have shown the results calculated by using 
 the DDM3Y force with  a density-dependence  of Kobos type \cite{Kobos1982}, which has been  
 successful for many calculations of nuclear rainbow scattering
 \cite{Ohkubo2004A,Ohkubo2007,Hamada2013,Hirabayashi2013,Ohkubo2014,Ohkubo2014B,Ohkubo2014C,Mackintosh2015,Ohkubo2015,Ohkubo2015B}. 
The $\chi^2$ values  at 74.25 MeV (157.7 MeV)   for the 
single channel (Fig.~1) and coupled channel (Fig.~2) calculations are 4.2  (8.4)   and 
2.4 (6.6), respectively, at  $\theta>$45$^\circ$ ($\theta>$30$^\circ$).
We show in Fig.~5 the calculated results using the zero-range M3Y force, which
 reproduce well the phases of the angular distribution of the Fraunhofer diffraction
 at forward angles in agreement with  the experiment at 157.7 MeV  as well as at 74.25 MeV. 
 The improvement of the phases  is due  a slightly shallower potential
 in the surface region. The used  potential parameters are  $N_R=$0.70, $W=18$ MeV and
 $a=0.85$ fm
for 157.7 MeV and   $N_R=$0.92, $W=15$ MeV and $a=1.0$ fm  for 74.25 MeV  with a fixed $R=5.5$ fm.
 That the primary bow is boosted dynamically by the coupling
 to the excited states of $^9$Be  does not change in the calculations using M3Y force.
The calculations using the DDM3Y force with zero-range also supports this conclusion.

\section{SUMMARY}

\par
To summarize,
we have presented 
  evidence for the existence  of a primary bow refracted  dynamically by  coupling
 to the  excited states of a weakly bound $^9$Be nucleus. 
The excited states play the role of a booster lens. 
   This is demonstrated by analyzing    $^{9}$Be+$^{16}$O rainbow scattering 
 using   the coupled
 channel method with  an extended double folding
(EDF) potential derived from the density-dependent effective two-body force with  
precise microscopic cluster wave functions for $^{9}$Be. 
The calculations reproduce the  Airy  structure in the experimental angular 
distributions  of $^{9}$Be+$^{16}$O rainbow scattering  well and clearly identify 
the existence of the Airy minimum $A1$ and the higher order Airy minimum $A2$.
About 10\% of the refraction is due to dynamical effects. 

\section{ACKOWLEDGEMENTS}
 \par
The authors thank the Yukawa Institute for Theoretical Physics, Kyoto University  for
 the hospitality extended  during a stay in February 2016 where this work 
 was completed.

\end{document}